# NONEQUILIBRIUM THERMODYNAMICS WITH THERMODYNAMIC PARAMETER OF LIFETIME OF SYSTEM. II. POSSIBILITIES OF INCREASING LIFETIME

V. V. Ryazanov

Institute for Nuclear Research, Kiev, Ukraine, e-mail: vryazan19@gmail.com

## Abstract

The thermodynamics is studied with the thermodynamic parameter of the lifetime, first-passage time, generalizing the equilibrium thermodynamics. Various ways of describing several stationary nonequilibrium states in the system are considered. The possibilities of increasing the lifetime of the system under external influences on it are investigated.

## 1. Introduction

Thermodynamics was considered in [1] in which the lifetime of the system, defined in the theory of random processes, was chosen as an additional thermodynamic variable, as the first-passage time, time the first zero level was reached by a random process for a macroscopic order parameter of the system, for example, the energy of the system or the number of particles in the system.

Such a description is phenomenological; it is related to the fact that the macroscopic parameters characterizing the behavior of the system become random under the influence of the environment. The term "lifetime" as applied to the considered circle of phenomena was used by R. L. Stratonovich in [2]. Synonyms of this term: first-passage time, the time of the first achievement of a certain level, the time to exit the area, the time of crossing the level, the busy period (in the queuing theory), etc.

To write the distributions from [1] containing a random value of the lifetime $\Gamma$, the dependence $\Gamma(z)$, the dependence of the random value of the lifetime $\Gamma(z)$ on the phase coordinates ($z=q_1,...,q_N, p_1,...,p_N$), where $p$ and $q$ are the momenta and coordinates of all particles in the system, are important.

The existence of such a dependence can be seen from the fact that the equation for the probability density of the distribution describing the behavior of value $\Gamma$, the Pontryagin equation [3], is conjugated by the kinetic equation (for example, the Fokker – Planck equation), which describes the probability distribution for energy. In [4] and [5] it is said that "Any function $B(z)$ of dynamic variables($z=q_1,...,q_N;p_1,...,p_N$) that is macroscopic in nature is a random internal thermodynamic parameter". The explicit form of the dependence $\Gamma$ is complex. The balance equation for $\Gamma$ is written in (26)-(27) [1].

The system in equilibrium is closed, there is no exchange with the environment, and the lifetime is difficult to determine. In non-stationary states, the mathematical definition of the time of the first achievement (the functional of the main random process) is not definided. To determine the lifetime, stationary nonequilibrium states, some "benchmark", reference points are needed.

We used in [1] the exponential distribution for the lifetime for the case of one stationary nonequilibrium state, and the Erlang distribution, the composition of the exponential distributions in the case of several stationary nonequilibrium states. In [6] it was shown that the exponential distribution for the lifetime appears as the limit distribution, in the limit of infinitely large times. But the study of finite lifetimes suggests the finiteness of the considered time intervals. It is necessary to use prelimit distributions.

The existence and finiteness of the lifetime value is ensured by the presence of stationary states, which physically corresponds to the existence of stationary structures. The complex functional and hierarchical relationships of real systems correspond to the same relationships between lifetimes, which determine evolutionary processes as a sequence of transitions between different classes of system states and on internal interactions in it.

In contrast to the traditional concept of time as something volatile, the lifetime is the result of the existence of stable stationary structures. It depends on external influences on the system and on internal interactions in it. We emphasize that the lifetime is an observable and measurable quantity. This is the same thermodynamic parameter as energy. Perhaps even more visual and obvious. It is contradictory in the traditional description of the lifetime that the diffusion approximation, most often used to determine the lifetime (for example, in the Kramers problem), characterizes small jumps, while transitions through the barrier are usually described by large jumps. Normal distribution should be replaced by gamma distribution.

It is shown in [1] that for the limit exponential distribution of the lifetime, any impact on one stationary nonequilibrium state only reduce its lifetime. In this paper it is shown that the impact can increase the average lifetime of the system in the case of the existence of several stationary nonequilibrium states in the system. The use of a complex Poisson distribution allows one to describe an ensemble of stationary nonequilibrium states. It is shown that in this case it is possible to increase the lifetime for one stationary nonequilibrium state with an exponential distribution of the lifetime.

We give some expressions obtained in [1] and used below.
The microscopic probability density in the extended phase space is

$$\rho(z;u,\Gamma)=exp\{-\beta u(z)-\gamma\Gamma(z)\}/Z(\beta,\gamma), \qquad (1)$$

where $u$ is the energy of the system,

$$Z(\beta,\gamma) = \int e^{-\beta u-\gamma\Gamma}dz = \iint du\, d\Gamma\, \omega(u,\Gamma)e^{-\beta u-\gamma\Gamma} \qquad (2)$$

is the partition function, $\beta$ and $\gamma$ are the Lagrange multipliers satisfying the following expressions for the averages:

$$<u>=-\partial lnZ/\partial\beta|_\gamma; \qquad <\Gamma>=-\partial lnZ/\partial\gamma|_\beta. \qquad (3)$$

These are relations (4)-(6) from [1].

The lifetime $\Gamma$ is defined as the random moment of the first achievement of a certain boundary (4), for example, of the zero level by a stochastic process that describes the macroscopic parameter $y(t)$, where $y(t)$ is the order parameter of the system, for example, its energy, number of particles,

$$\Gamma_x = \inf\{t: y(t)=0\}, \quad y(0)=x>0 \tag{4}$$

(the relation (1) from [1]). The lifetime is the time period till the moment of the first (random) reattachment of a certain level (for example, zero level) by a random process $y(t)$ for the macroscopic parameter (for example, energy or particle number).

We introduce the entropy $S_\Gamma$ corresponding to distribution (1) by the relation

$$S_\Gamma = -\langle \ln\rho(z;u,\Gamma)\rangle = \beta\langle u\rangle + \gamma\langle\Gamma\rangle + \ln Z(\beta,\gamma); \quad dS_\Gamma = \beta d\langle u\rangle + \gamma d\langle\Gamma\rangle. \quad S_\Gamma = s_{eq}-\Delta. \tag{5}$$

(the relation (7) from[1]). In distribution (1)-(2), containing the lifetime as a thermodynamic parameter, the joint probability (1) for the quantities $u$ and $\Gamma$ is

$$P(u,\Gamma) = \frac{e^{-\beta u - \gamma \Gamma}\omega(u,\Gamma)}{Z(\beta,\gamma)} \tag{6}$$

(the relation (10) from[1]). Integrating (6) over $\Gamma$, we obtain a distribution of the form

$$P(u) = \int P(u,\Gamma)d\Gamma = \frac{e^{-\beta u}}{Z(\beta,\gamma)}\int_0^\infty e^{-\gamma\Gamma}\omega(u,\Gamma)d\Gamma. \tag{7}$$

The factor $\omega(u,\Gamma)$ is the joint probability for $u$ and $\Gamma$, considered as the stationary probability of this process. We rewrite the value $\omega(u,\Gamma)$ in the form

$$\omega(u,\Gamma) = \omega(u)\omega_1(u,\Gamma) = \omega(u)\sum_{k=1}^{n} R_k f_{1k}(\Gamma,u) \tag{8}$$

(the relation (12) from [1]; $R_k$ is the probability that the system is in the $k$-th class of states, $f_{1k}(\Gamma,u)$ is the density of the distribution of the lifetime $\Gamma$ in this class of states). To obtain an explicit form of the distribution $f_{1k}(\Gamma,u)$ in (8), we use the general results of the mathematical theory of phase enlargement of complex systems, from which the following exponential form of the density of the distribution of the lifetime for an enlarged random process follows:

$$p(\Gamma<y) = \Gamma_0^{-1}\exp\{-y/\Gamma_0\} \tag{9}$$

(the relation (13) from [1]) for one class of stable states, and Erlang distribution density

$$p(\Gamma<y) = \sum_{k=1}^{n} R_k \Gamma_{0k}^{-1}\exp\{-y/\Gamma_{0k}\}; \quad \sum_{k=1}^{n} R_k = 1 \tag{10}$$

(the relation (14) from [1]) in the case of several ($n$) classes of ergodic stable states, $\langle\Gamma_{\gamma k}\rangle$ is the average lifetime in the $k$-th class of metastable states with a perturbation $\gamma=\gamma_k$, $\Gamma_{0k}=\langle\Gamma_{0k}\rangle$ is the average lifetime in the $k$-th class of metastable states without a perturbation $\gamma=\gamma_k=0$, in some basic stationary state. Ratios are also used:

$$a_\beta q^2/2 = \ln(1+x), \tag{11}$$

(the relation (45) from[1]),

$$\gamma = \frac{1}{\Gamma_0}(e^{a_\beta q^2/2}-1), \quad 1+x = e^{a_\beta q^2/2}, \quad \langle\Gamma\rangle = \frac{\Gamma_0}{1+x} = \Gamma_0 e^{-a_\beta q^2/2}, \tag{12}$$

(the relation (46) from[1]),

$$a_\beta q \gamma \frac{dq}{d\gamma} = \frac{x}{1+x}, \quad \frac{1}{2}q^2\beta\frac{da_\beta}{d\beta} = -\frac{x}{1+x} = -\gamma\langle\Gamma\rangle, \tag{13}$$

(the relation (47) from[1]).

## 2. Possibilities to increase the average lifetime

Note that in [1] such quantities are introduced as the random value of the lifetime $\Gamma$, the nonequilibrium parameter $\gamma$ conjugate to this quantity, which in [1] is associated with a change in the entropy $-\Delta = s - s_{eq}$ ($s_{eq}$ is equilibrium entropy), the nonequilibrium part of the energy $<u_\gamma> = -\dfrac{\partial \ln(Z/Z_\beta)}{\partial \beta}\Big|_\gamma$, where $Z$ from (2), $Z_\beta$ is the equilibrium partition function. New values correspond to new relationships between them. So in [1] the relation are obtained

$$\beta <u_\gamma> + \gamma \langle \Gamma \rangle = 0 \qquad (14)$$

If the distribution (1) are written as $\rho \sim \exp(-L)$, then with $\bar{u} = u_\beta + u_\gamma$, $\langle L \rangle = \beta u_\beta$, which corresponds to the Gibbs distribution. But the normalization of the non-averaged distribution is different, and expression (5) takes the form $\ln Z_\gamma = -\Delta$.

The relations obtained in [1] are written for deviations from the equilibrium state of one stationary nonequilibrium state. If we consider several stationary nonequilibrium states and transitions between them, then there are opportunities to increase the average lifetime of the system. Consider the case of two stationary nonequilibrium states, two "potential wells" using expressions (7)-(10) with $n=2$. Choosing, as in (10), the exponential distributions for the lifetime, we obtain

$$Z = Z_\beta Z_\gamma, \quad Z_\gamma = \sum_{i=1}^{2} P_i m_i, \quad m_i = \frac{1}{1+x_i}, \quad x_i = \gamma_i \Gamma_{0i}; \quad i=1,2; \qquad (15)$$

$$\bar{u} = -\frac{d \ln Z}{d\beta} = u_\beta + \frac{1}{Z_\gamma} \sum_{i=1}^{2} (P_i m_i^2 \gamma_i \frac{\partial \Gamma_{0i}}{\partial \beta} - m_i \frac{\partial P_i}{\partial \beta}), \quad u_\beta = -\frac{d \ln Z_\beta}{d\beta}, \quad Z_\beta = \int e^{-\beta u} \omega(u) du, \qquad (16)$$

$$\bar{\Gamma} = -\frac{d \ln Z}{d\gamma} = \frac{1}{Z_\gamma} \sum_{i=1}^{2} (P_i m_i^2 \Gamma_{0i} - m_i \frac{\partial P_i}{\partial \gamma}) \qquad (17)$$

Using relations (5) for the desired quantities $m_1$ and $m_2$, we write two equations:

$$s_i = s_\beta - \Delta_i = \gamma_i \bar{\Gamma} + \beta \bar{u} + \ln Z, \quad \Delta_i = a_\beta q_i^2 / 2, \quad \Delta_i = \gamma_i \bar{\Gamma} + \beta \bar{u}_\gamma + \ln Z_\gamma, \quad i=1,2, \quad \bar{u}_\gamma = \bar{u} - \bar{u}_\beta. \qquad (18)$$

We assume that $\beta$ is the same in both states; only parameter $\gamma$ changes. Suppose that $\gamma$ takes the value $\gamma_1$ in state 1 and $\gamma_2$ in state 2. We choose the distribution $R_i, i=1,2$ from (10) in the form

$$R_i = \frac{e^{-s_i}}{e^{-s_1} + e^{-s_2}}, \quad s_i = s_\beta - \Delta_i, \quad \Delta_i = a_\beta q_i^2 / 2, \quad a_\beta = \frac{\tau}{\rho \lambda T^2}. \qquad (19)$$

The solution of equations (17) - (18) has the form:

$$\gamma_1 \Gamma_{01} = \frac{D_1 B_2 + A_1 D_2}{B_1 B_2 - A_1 A_2}, \quad \gamma_2 \Gamma_{02} = \frac{D_2 B_1 + A_2 D_1}{B_1 B_2 - A_1 A_2}, \qquad (20)$$

where

$$D_2 = Z_\gamma [\ln 2 - \ln(1 + e^{\Delta_2 - \Delta_1})], \quad Z_\gamma = \frac{2}{(e^{\Delta_2} + e^{\Delta_1})}, \quad D_1 = D_2 + Z_\gamma (\Delta_2 - \Delta_1),$$

$$B_2 = A_2 \frac{\Gamma_{01}}{\Gamma_{02}} - D_2, \quad B_1 = A_1 \frac{\Gamma_{02}}{\Gamma_{01}} - D_1, \quad A_1 = \frac{2}{K} \frac{\Gamma_{01}}{\Gamma_{02}} e^{\Delta_1}, \quad A_2 = \frac{2}{K} \frac{\Gamma_{02}}{\Gamma_{01}} e^{\Delta_2}, \quad K = (e^{\Delta_1} + e^{\Delta_2})^2.$$

In order for the average lifetime of the system to increase, the inequality must hold

$$\bar{\Gamma} > \bar{\Gamma}_0, \quad \bar{\Gamma}_0 = \bar{\Gamma}_{\gamma=0} = \Gamma_{01} R_1 + \Gamma_{02} R_2. \qquad (21)$$

Substituting expressions (19) into (15)-(18), (21), we obtain, using expressions (13), the exact inequality necessary to satisfy condition (21)

$D_2\{Z_\gamma(\Delta_2-\Delta_1)[(2\Gamma_{02}/K\Gamma_{01})exp\{\Delta_2\}-D_2]+D_2[(2\Gamma_{02}/K\Gamma_{01})exp\{\Delta_2\}-D_2+(2/K)exp\{\Delta_1\}]\}>$

$>[Z_\gamma(\Delta_2-\Delta_1)+D_2]\{Z_\gamma(\Delta_2-\Delta_1)[(2/K)exp\{\Delta_2\}-D_2]+D_2[(2\Gamma_{01}/K\Gamma_{02})exp\{\Delta_1\}-D_2+(2/K)exp\{\Delta_2\}]\}$. (22)

Assuming that $\Delta_2-\Delta_1$ is small, in the linear approximation in $\Delta_2-\Delta_1$, the inequality necessary to fulfill condition (64) takes the form

$$-(\frac{3}{2}+\frac{1}{2}\frac{\Gamma_{02}}{\Gamma_{01}})[\sqrt{1+\frac{(\Gamma_{02}/\Gamma_{01})^2-1}{(\frac{3}{2}+\frac{1}{2}\frac{\Gamma_{02}}{\Gamma_{01}})^2}}+1] < \Delta_2-\Delta_1 < (\frac{3}{2}+\frac{1}{2}\frac{\Gamma_{02}}{\Gamma_{01}})[\sqrt{1+\frac{(\Gamma_{02}/\Gamma_{01})^2-1}{(\frac{3}{2}+\frac{1}{2}\frac{\Gamma_{02}}{\Gamma_{01}})^2}}-1].$$

It was noted in [7], [8] that there are many possibilities for choosing the distribution of the lifetime from (8), which was chosen in (9) as an exponential distribution. Instead of the exponential distribution for the lifetime, for example, you can specify the gamma distribution of the form

$$f(x) = \frac{1}{\Gamma(c)}\frac{1}{b^c}x^{c-1}e^{-x/b}$$

($\Gamma(c)$ is the gamma function). Such a distribution for the lifetime is associated in [7], [8] with obtaining superstatistics and the Tsallis distribution from distributions with the lifetime. The relations obtained in [1] change. So, instead of expression (11), (45) from [1], $a_\beta q^2/2 = \ln(1+x)$ expression is written $a_\beta q^2/2 = c\ln(1+x/c)$, where $c$ is the gamma distribution parameter. Relations (21)-(22) written down not for the exponential, but for the case of the gamma distribution of the lifetime in both stationary nonequilibrium states, takes the form

$D_2\{Z_\gamma(\Delta_2-\Delta_1)[(2\Gamma_{02}/K\Gamma_{01})exp\{\Delta_2\}c_1-D_2]+D_2[2\Gamma_{02}/K\Gamma_{01}))exp\{\Delta_2\}c_1-D_2+(2/K)exp\{\Delta_1\}c_1]\}>$

$>[Z_\gamma(\Delta_2-\Delta_1)+D_2]\{Z_\gamma(\Delta_2-\Delta_1)[(2/K)exp\{\Delta_2\}c_2-D_2]+D_2[(2\Gamma_{01}/K\Gamma_{02})exp\{\Delta_1\}c_2-D_2+(2/K)exp\{\Delta_2\}c_2]\}$,

where $c_i$ are the gamma distribution parameters in the $i$-th potential well ($i = 1,2$), the expressions for $D_2$ are written in (20), and the expressions for $B_i$ take the form

$$B_2 = A_2\frac{\Gamma_{01}}{\Gamma_{02}}-\frac{D_2}{c_2}, \quad B_1 = A_1\frac{\Gamma_{02}}{\Gamma_{01}}-\frac{D_1}{c_1}.$$

The linear approximation also changes, taking the form

$$-[1+\frac{1}{2}(c_2+c_1\frac{\Gamma_{02}}{\Gamma_{01}})][\sqrt{1+\frac{(c_2+c_1\frac{\Gamma_{02}}{\Gamma_{01}})(\frac{\Gamma_{02}}{\Gamma_{01}}-1)}{[1+\frac{1}{2}(c_2+c_1\frac{\Gamma_{02}}{\Gamma_{01}})]^2}}+1] < \Delta_2-\Delta_1 <$$

$$< [1+\frac{1}{2}(c_2+c_1\frac{\Gamma_{02}}{\Gamma_{01}})][\sqrt{1+\frac{(c_2+c_1\frac{\Gamma_{02}}{\Gamma_{01}})(\frac{\Gamma_{02}}{\Gamma_{01}}-1)}{[1+\frac{1}{2}(c_2+c_1\frac{\Gamma_{02}}{\Gamma_{01}})]^2}}-1].$$

If set independent, $x_i$, $1+x_i=e^{\Delta_i}$ then $\overline{\Gamma}=\frac{\Gamma_{01}+\Gamma_{02}}{e^{\Delta_1}+e^{\Delta_2}}$, and $\overline{\Gamma} \leq \Gamma_0$, no increase in average lifetime. But $x_i$ are connected, you cannot set them independently.

The intermediate case was considered above, when the approximation $1+x_i = e^{\Delta_i}$ was not used everywhere. In the general case, it is necessary to solve the system of equations for $x_i$. If we solve the equations $\ln Z_\gamma + \Delta_i + \gamma_i\overline{\Gamma} + \beta\overline{u}_\gamma = 0$, $i=1,2$, using everywhere, as an approximation, conditions $1+x_i=e^{\Delta_i}$, we get approximate expressions

$$Z_\gamma = \frac{2}{e^{\Delta_1}+e^{\Delta_2}}, \qquad Z_\gamma \bar{\Gamma} = \frac{2(\Gamma_{01}+\Gamma_{02})}{K}, \qquad K=(e^{\Delta_1}+e^{\Delta_2})^2, \qquad Z_\gamma \beta u_\gamma = \frac{2}{e^{\Delta_1}+e^{\Delta_2}}(\frac{2}{e^{\Delta_1}+e^{\Delta_2}}-1).$$

We find

$$-\gamma_i = \frac{Z_\gamma(\Delta_i + \ln Z_\gamma + \beta u_\gamma)}{(\Gamma_{01}+\Gamma_{02})\frac{2}{K}},$$

And the condition $\bar{\Gamma} \geq \Gamma_0$, $\frac{\Gamma_{01}e^{\Delta_1}}{1+x_1} + \frac{\Gamma_{02}e^{\Delta_2}}{1+x_2} \geq \Gamma_{01}e^{\Delta_1} + \Gamma_{02}e^{\Delta_2}$, takes the form

$$\frac{\Gamma^2_{01}\Delta_1 + \Gamma^2_{02}\Delta_2}{\Gamma^2_{01}+\Gamma^2_{02}} > 1 - \frac{2}{e^{\Delta_1}+e^{\Delta_2}} - \ln\frac{2}{e^{\Delta_1}+e^{\Delta_2}} = 1 - Z_\gamma - \ln Z_\gamma. \tag{23}$$

If we do not use the independence approximation for $x_i$, then the full expressions are written as

$$Z_\gamma = \frac{1}{e^{\Delta_1}+e^{\Delta_2}}[\frac{e^{\Delta_1}}{1+x_1} + \frac{e^{\Delta_2}}{1+x_2}],$$

$$Z_\gamma \bar{\Gamma} = \frac{1}{K}[\frac{\Gamma_{01}e^{2\Delta_1}}{1+x_1} + \frac{\Gamma_{01}e^{2\Delta_2}}{1+x_2} + e^{\Delta_1+\Delta_2}(\frac{\Gamma_{01}x_1}{(1+x_1)^2} + \frac{\Gamma_{02}x_2}{(1+x_2)^2} + \frac{\Gamma_{01}}{(1+x_1)} + \frac{\Gamma_{02}}{(1+x_2)})],$$

$$Z_\gamma \beta u_\gamma = \frac{1}{K}[-\frac{x_1 e^{2\Delta_1}}{1+x_1} - \frac{x_2 e^{2\Delta_2}}{1+x_2} - e^{\Delta_1+\Delta_2}(\frac{x_1+x_2}{(1+x_1)(1+x_2)} + \frac{x_1^2}{(1+x_1)^2} + \frac{x_2^2}{(1+x_2)^2})].$$

And the equations for $x_i$ are written as

$$Z_\gamma(\ln Z_\gamma + \Delta_1) + \frac{1}{K}(x_1\frac{\Gamma_{02}}{\Gamma_{01}} - x_2)\frac{e^{\Delta_2}}{1+x_2}[e^{\Delta_2} + e^{\Delta_1}(\frac{x_2}{1+x_2} + \frac{1}{1+x_1})] = 0,$$

$$Z_\gamma(\ln Z_\gamma + \Delta_2) + \frac{1}{K}(x_2\frac{\Gamma_{01}}{\Gamma_{02}} - x_1)\frac{e^{\Delta_1}}{1+x_1}[e^{\Delta_1} + e^{\Delta_2}(\frac{x_1}{1+x_1} + \frac{1}{1+x_2})] = 0.$$

If we solve these equations in a linear approximation in $x_i$, then

$$Z_\gamma \approx 1 - \frac{x_1 e^{\Delta_1} + x_2 e^{\Delta_2}}{e^{\Delta_1}+e^{\Delta_2}}, \quad x_i \approx \frac{(e^{\Delta_1}+e^{\Delta_2})a_i}{M}, \quad a_1 = e^{\Delta_1}[\Delta_1(\frac{\Gamma_{01}}{\Gamma_{02}}-\Delta_2)] - e^{\Delta_2}(\Delta_1+\Delta_1\Delta_2-2\Delta_2),$$

$$a_2 = e^{\Delta_2}[\Delta_2(\frac{\Gamma_{02}}{\Gamma_{01}}-\Delta_1)] - e^{\Delta_1}(\Delta_2+\Delta_1\Delta_2-2\Delta_1),$$

$$M = e^{2\Delta_1}[\frac{\Gamma_{01}}{\Gamma_{02}} + \Delta_2 + 2\Delta_1(\Delta_2-2)] + e^{2\Delta_2}[\frac{\Gamma_{02}}{\Gamma_{01}} + \Delta_1 + 2\Delta_2(\Delta_1-2)] +$$

$$+ e^{\Delta_1+\Delta_2}(2+4\Delta_1\Delta_2 - 2\Delta_2 - 2\Delta_1 - \Delta_2\frac{\Gamma_{02}}{\Gamma_{01}} - \Delta_1\frac{\Gamma_{01}}{\Gamma_{02}}).$$

The growth condition of the average lifetime in the linear approximation in $x_i$ takes the form

$$\Gamma_{01}e^{\Delta_1}\{\frac{M+a_2\Delta_\Sigma}{\Delta_\Sigma(M+a_1 e^{\Delta_2} + a_2 e^{\Delta_1})}[e^{\Delta_1} + e^{\Delta_2}(1+\frac{(a_1-a_2)\Delta_\Sigma M}{(M+a_1\Delta_\Sigma)(M+a_2\Delta_\Sigma)})]-1\} +$$

$$+\Gamma_{02}e^{\Delta_2}\{\frac{M+a_1\Delta_\Sigma}{\Delta_\Sigma(M+a_1 e^{\Delta_2} + a_2 e^{\Delta_1})}[e^{\Delta_2} + e^{\Delta_1}(1-\frac{(a_1-a_2)\Delta_\Sigma M}{(M+a_1\Delta_\Sigma)(M+a_2\Delta_\Sigma)})]-1\} > 0;$$

$$\Delta_\Sigma = e^{\Delta_1}+e^{\Delta_2}. \tag{24}$$

If in this relation we consider the linear approximation and for $\Delta_1, \Delta_2$, then the condition for the growth of the average lifetime takes the form

$$p_0 + p_1\Delta_1 + p_2\Delta_2 > 0, \; p_0 = \Gamma_{01}+\Gamma_{02};$$

$$p_1 = 2\Gamma_{01} + \Gamma_{02} + (\Gamma_{01} - \Gamma_{02})\frac{\alpha_1}{\alpha}(\frac{\Gamma_{01}}{\Gamma_{02}} - 3) + \frac{2}{\alpha_1}(5\Gamma_{01} - 9\Gamma_{02} - \frac{\Gamma_{02}^2}{\Gamma_{01}} - 3\frac{\Gamma_{01}^2}{\Gamma_{02}}); \qquad (25)$$

$$p_2 = 2\Gamma_{02} + \Gamma_{01} + (\Gamma_{01} - \Gamma_{02})\frac{\alpha_1}{\alpha}(3 - \frac{\Gamma_{02}}{\Gamma_{01}}) + \frac{2}{\alpha_1}(5\Gamma_{02} - 9\Gamma_{01} - 3\frac{\Gamma_{02}^2}{\Gamma_{01}} - \frac{\Gamma_{01}^2}{\Gamma_{02}});$$

$$\alpha_1 = 2(2 + \frac{\Gamma_{01}}{\Gamma_{02}} + \frac{\Gamma_{02}}{\Gamma_{01}}) \qquad \alpha = 6 + 2(\frac{\Gamma_{01}}{\Gamma_{02}} + \frac{\Gamma_{02}}{\Gamma_{01}}) + (\frac{\Gamma_{02}}{\Gamma_{01}})^2 + (\frac{\Gamma_{01}}{\Gamma_{02}})^2.$$

## 3. Complex Poisson distribution for lifetime and a random number of potential wells.

In addition to using the Erlang distribution (10), there are other possibilities for describing several stationary nonequilibrium states. Instead of the Erlang distribution (10), we use the generalized Poisson distribution with the Laplace transform

$$\ln \varphi(\gamma) = -a(1 - p(\gamma)). \qquad (26)$$

The meaning of the notation from (26) is explained below. Growth conditions for the average lifetime of a system (for example, of the form (23)-(25)) in this case become simpler. They do not include the probabilities $R_i$ from relations (8)-(10). One complex state with a statistical structure is considered. An important role is played by the ratio of the change in entropy to the average number of states a. The potential of the system described by distribution (26) is shown in Fig. 1, taken from [9].

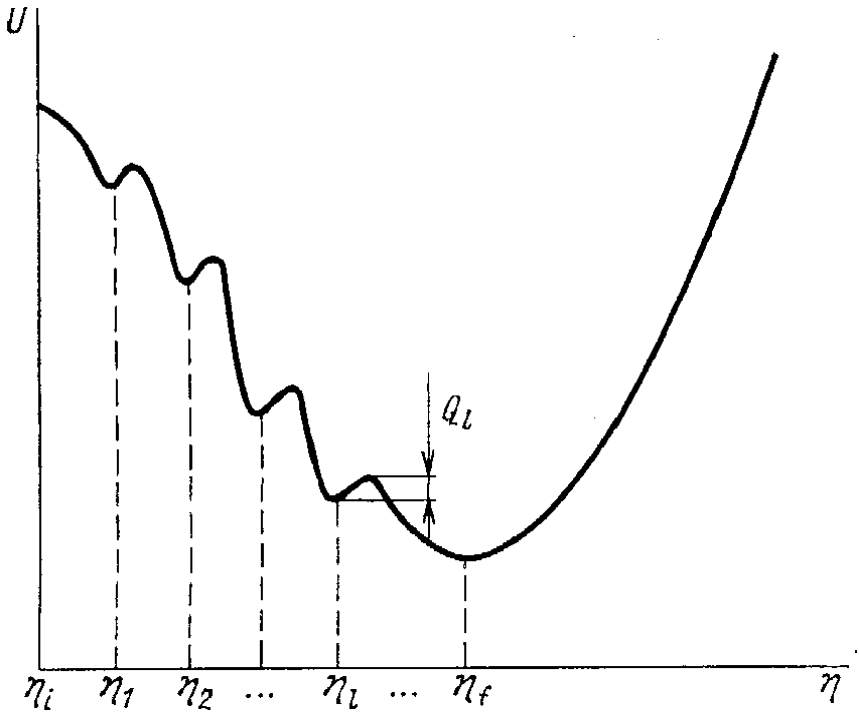

• Fig. 1 [9]. Schematic dependence of the thermodynamic potential $U$ on the order parameter $\eta$ for the case when the phase space of the system is divided into isolated regions, in each of which a metastable thermodynamic state is established.

Exponential distributions for the lifetime were considered in [1]. Now we consider a complex distribution for the lifetime corresponding to Figure 1 and a random number of potential wells. The generating function of Poisson distribution is $\varphi(z) = e^{-a(1-z)}$.

If the argument $z$ in this expression is replaced by the Laplace transform of the exponential distribution $p(\gamma) = \dfrac{1}{1+\gamma\Gamma_{01}}$, $x = \gamma\Gamma_{01}$, then we obtain the Laplace transform of the form (26) of a distribution describing the sum $\sum_{k=1}^{\nu} \zeta_k$ of independent random variables with the Laplace transform $p_\eta(\gamma) = \varphi(p(\gamma))$ equally distributed with the exponential distribution, where $\nu$ is an integer non-negative random variable with generating function $\varphi(z) = e^{-a(1-z)}$.

This is the sum of a random number of potential wells with an exponential distribution of the system lifetime in each potential well; the value $a$ represents the average number of potential wells in this sum. If the statistical sum is written as $Z(\beta,\gamma) = Z(\beta)Z(\gamma)$, then the part of the partition function $Z(\gamma) = \int e^{-\gamma\Gamma} p(\Gamma) d\Gamma$, which previously had the form $p(\gamma) = \dfrac{1}{1+\gamma\Gamma_{01}}$, $x = \gamma\Gamma_{01}$, is now equal to $Z(\gamma) = \int e^{-\gamma\Gamma} p(\Gamma) d\Gamma = e^{-a\frac{x}{1+x}}$;

$$\ln Z_\gamma = -\frac{ax}{1+x}, \quad x = \gamma\Gamma_{01}, \qquad (27)$$

$$\overline{\Gamma} = -\frac{\partial \ln Z_\gamma}{\partial \gamma} = \frac{a\Gamma_{01}}{(1+x)^2}, \quad s = \beta\overline{u} + \gamma\overline{\Gamma} + \ln Z = \beta u_\beta + \ln Z_\beta - \Delta_1, \quad \Delta_1 = a_\beta q_2/2, \quad \overline{u} = u_\beta + u_\gamma. \qquad (28)$$

Assume that the parameter $a$ depends on $\beta$. Then

$$u_\gamma = \frac{x}{1+x}\frac{\partial a}{\partial \beta} + \frac{\partial x}{\partial \beta}\frac{a}{(1+x)^2}. \qquad (29)$$

If to $\Gamma_{01}$ use the expression $\beta\dfrac{d\overline{\Gamma}}{d\beta} = -\dfrac{\overline{\Gamma}}{1+x}$, $\beta\dfrac{d\Gamma_0}{d\beta} = -\Gamma_0$, $\dfrac{d\overline{\Gamma}}{d\gamma} = -\langle\Gamma\rangle^2$ ((49) from [1]), then

$$\frac{\partial \Gamma_{01}}{\partial \beta} = -\frac{\Gamma_{01}}{\beta}, \quad \frac{\partial x}{\partial \beta} = -\frac{x}{\beta}.$$

The ratio is fair (instead of (14))

$$\beta \langle u_\gamma \rangle + \gamma\langle\Gamma\rangle = \frac{x}{1+x}\frac{\partial a}{\partial \beta}.$$

Substituting these expressions into the relation for entropy

$$\ln Z_\gamma = -\frac{ax}{1+x} = -\Delta - \frac{\partial a}{\partial \beta}\frac{x}{1+x}, \quad x = \gamma\Gamma_{01},$$

we get that

$$\Delta = (a - \frac{\partial a}{\partial \beta})\frac{x}{1+x}, \quad x = \frac{\Delta a_1^{-1}}{1 - \Delta a_1^{-1}}, \quad a_1 = a - \frac{\partial a}{\partial \beta},$$

$$\overline{\Gamma} = a\Gamma_{01}(1 - \Delta a_1^{-1})^2. \qquad (30)$$

And the condition for the growth of the average lifetime takes the form

$$\overline{\Gamma} > \Gamma_0 = a\Gamma_{01}, \quad \Delta/a_1 > 2. \qquad (31)$$

The condition for the growth of the average lifetime $\bar{\Gamma} > \bar{\Gamma}_{\gamma=0} = a\Gamma_{01}$ (31) is also satisfied at $\Delta > 0$, $a_1 < 0$, $x < 0$, $a < \beta \dfrac{da}{d\beta}$. Wherein $\Delta_1 = a_1(\beta)(1 - e^{-\Delta_{10}})$, $\Delta_{10} = \ln(1+x)$, $\Delta = -(s - s_{eq})$.

When $a_1 > 0$, $x > 0$, $a > \beta \dfrac{da}{d\beta}$, $\Delta/a_1 < 2$, there is a decrease in average lifetime;.

If we assume that the parameter $a$ depends on $\gamma$, then $\bar{\Gamma} = \dfrac{\Gamma_{01} a}{(1+x)^2} + \dfrac{x}{1+x}\dfrac{\partial a}{\partial \gamma}$. Substituting these expressions into the relation for entropy, we obtain the same expressions (30)-(31), but with

$$a_1 = a - \beta \frac{da}{d\beta} - \gamma \frac{da}{d\gamma} .$$

The same relation is obtained from the thermodynamic relation ((23) from [1])

$$\gamma = \frac{1}{\dfrac{\partial \bar{\Gamma}}{\partial \beta}\bigg|_\gamma}(\theta^{-1}\frac{\partial \bar{u}_\beta}{\partial \beta}\bigg|_\gamma - \beta \frac{\partial \bar{u}}{\partial \beta}\bigg|_\gamma) .$$

The homogeneous case is considered above, when the quantities $\beta, \gamma, \Gamma_0$ are the same in all $a$ states. But the value of $a$ can also be considered as an independent variable, like $\beta, \gamma, \Gamma_0$. You can determine $-\partial \ln Z_\gamma / \partial a = x/1+x$ - an analog of the pressure $P$, if $a$ is an analog of the volume, then, for example, the Maxwell relations are satisfied: $\partial P/\partial \beta = \partial \bar{u}/\partial a$, $\partial P/\partial \gamma = \partial \bar{\Gamma}/\partial a$, and the value of $a$ is independent of $\beta, \gamma, \Gamma_0$ the parameter.

The equality of all parameters corresponds to equilibrium, and we consider the nonequilibrium case. Therefore, we consider a situation where the parameters in different states differ: $\beta \to \beta_i, \gamma \to \gamma_i, \Gamma_{01} \to \Gamma_{0i}, \Delta_1 \to \Delta_{1i}, i=1,...,a$. With this approach, unlike the equal parameters, the growth or decrease of $\bar{\Gamma}$ is not affected by the values of $a$ and $da/d\beta$, but by the differences between the states. The quantity $\Gamma_{0i}$ determines the depth of the $i$-th potential well.

Put $a$=const, independent of $\beta, \gamma, \Gamma_0$. In the expression $-\ln Z_\gamma = a(1 - \varphi(\gamma))$ we replace $\varphi(\gamma) = 1/(1+x)$, $x = \gamma \Gamma_{01}$ with the Laplace transform of the form $\dfrac{1}{a}\sum_{i=1}^{a}\varphi_i(\gamma_i)$, where $\varphi_i(\gamma_i)$ is the Laplace transform of the distribution of the lifetime for the $i$-th state with parameters $\gamma_i$, $\Gamma_{0i}$.

For example, for the limiting exponential distribution in all states, $\varphi_\kappa(\gamma_\kappa) = \dfrac{1}{1+x_\kappa}$, $x_\kappa = \gamma_\kappa \Gamma_{0\kappa}$ for each of $a$ states, we select the exponential distribution for the lifetime, but the parameters of this distribution differ in all states, as well as the parameter of the Laplace transform $\gamma_i$, the value conjugate with the random lifetime of the $i$-th state.

Then

$$-\ln Z_\gamma = a(1 - \frac{1}{a}\sum_{\kappa=1}^{a}\frac{1}{1+x_\kappa}). \qquad (32)$$

If we assume that the operations $\dfrac{\partial}{\partial \gamma_i} \to \dfrac{\delta}{\delta \gamma_i}$, $\dfrac{\partial}{\partial \beta_i} \to \dfrac{\delta}{\delta \beta_i}$ act on the $i$-th state, the $i$-th cell of the system, then $\bar{\Gamma}_i = -\delta \ln Z_\gamma / \delta \gamma_i = \Gamma_{0i}/(1+x_i)^2$, $\bar{\Gamma} = \sum_{i=1}^{a}\bar{\Gamma}_i$.

In the homogeneous case, this expression for $x_i = x$, $\Gamma_{01} = \Gamma_{0Ii}$ coincides with the expression (28) obtained earlier. So it is for energy, when the parameter $a$ does not depend on $\beta$, $\bar{u}_i = -\delta \ln Z_\gamma / \delta \beta_i = x_i / \beta_i (1+x_i)^2$. We use expressions $\partial x_i / \partial \beta_i = -x_i / \beta_i$, $\bar{u}_\gamma = \sum_{i=1}^{a} \bar{u}_{\gamma i}$ that coincide with the homogeneous case with the same parameters.

In relation (5): $-\ln Z_\gamma = \beta \bar{u}_\gamma + \gamma \bar{\Gamma} + \Delta_1$ we replace $\beta \bar{u}_\gamma$ by $\sum \beta_i \bar{u}_{\gamma i} = \sum_{i=1}^{a} \frac{x_i}{(1+x_i)^2}$; this expression goes over into $-ax/(1+x)^2$ in the homogeneous case that obtained previously in (30). We replace the value $\gamma \bar{\Gamma}$ by $\sum_{i=1}^{a} \gamma_i \bar{\Gamma}_i = \sum_{i=1}^{a} \frac{x_i}{(1+x_i)^2}$, which also goes into expression (28) in the homogeneous case. Then for $\Delta = \sum_{i=1}^{a} \Delta_i$,

$$-\ln Z_\gamma = \sum \beta_i \bar{u}_{\gamma i} + \sum \gamma_i \bar{\Gamma}_i + \frac{1}{a} \sum \Delta_{1i}$$

- this expression in the homogeneous case goes over into (5). Then

$$-\ln Z_\gamma = a - \sum \frac{1}{(1+x_i)} = \frac{1}{a} \sum \Delta_{1i}, \quad \beta_j \frac{\partial \bar{\Gamma}}{\partial \beta_j} = -\frac{\Gamma_{0j}(1-x_j)}{(1+x_j)^3}.$$

From the relations

$$-\partial \ln Z_\gamma / \partial \gamma_i = \bar{\Gamma}_i = \beta_i \partial \bar{u}_{\gamma i} / \partial \gamma_i + \bar{\Gamma}_i + \gamma_i \partial \bar{\Gamma}_i / \partial \gamma_i + \frac{1}{a} \partial \Delta_{1i} / \partial \gamma_i, \quad \partial \bar{u}_{\gamma i} / \partial \gamma_i = \partial \bar{\Gamma}_i / \partial \beta_i,$$

$\gamma_i \partial \bar{\Gamma}_i / \partial \gamma_i = -2 x_i \Gamma_{0i} / (1+x_i)^3$, $\beta_i \partial \bar{\Gamma}_i / \partial \beta_i = -\frac{1}{a} \frac{\partial \Delta_{1i}}{\partial \gamma_i} + \frac{2 x_i \Gamma_{0i}}{(1+x_i)^3} = -\frac{\Gamma_{0i}(1-x_i)}{(1+x_i)^3}$, $\frac{1}{a} \frac{\partial \Delta_{1i}}{\partial \gamma_i} = \frac{\Gamma_{0i}}{(1+x_i)^2}$,

$\Delta_{1i} = -a/(1+x_i) + c_1$, $c_1 = a$, so that $\Delta_{1|x=0} = 0$, we get

$\Delta_{1i} = a x_i / (1+x_i)$, $x_i = (\Delta_{1i}/a)/(1-(\Delta_{1i}/a))$, and as in (30),

$$\bar{\Gamma}_i = \Gamma_{0i}(1 - \frac{\Delta_{1i}}{a})^2. \tag{33}$$

We get that as in (31)

$$\frac{\bar{\Gamma}_i}{\Gamma_{0i}} \leq 1 \text{ at } (\Delta_{1i}/a) \leq 2, \quad \frac{\bar{\Gamma}_i}{\Gamma_{0i}} > 1 \text{ at } (\Delta_{1i}/a) > 2.$$

Note that this result for $a=1$ does not agree with result [1]. For one stationary nonequilibrium state it becomes possible to increase the lifetime. This is a consequence of the fact that in [1] we used the exponential distribution for the lifetime, and here is the complex Poisson distribution. In the latter case not one stationary nonequilibrium state is considered, but an ensemble of states. This approach seems preferable. For the full average lifetime of the entire system, we obtain

$$\bar{\Gamma} = \sum_{i=1}^{a} \bar{\Gamma}_i = \sum_{i=1}^{a} \Gamma_{0i}(1-\frac{\Delta_{1i}}{a})^2, \quad \bar{\Gamma} = \sum_{i=1}^{a} \bar{\Gamma}_i > \Gamma_0 = \sum_{i=1}^{a} \Gamma_{0i} \text{ at } \bar{\Gamma}/\Gamma_0 = \sum_{i=1}^{a} \Gamma_{0i}(1-\frac{\Delta_{1i}}{a})^2 / \sum_{i=1}^{a} \Gamma_{0i} > 1.$$

Different states can make different contributions to the total average lifetime of the entire system. Thus, an increase or decrease in the average lifetime from the impacts depends on the ratio of the change in entropy $\Delta_{1i}$ to the average number of states $a$.

In the general case, an increase in the average lifetime under impacts is possible either under the condition $(\Delta_{1i}/a) > 2$, $\Delta_{1i} > 0$ or under the condition $\Delta_{1i} \to \Delta s < 0$ when flows of negative entropy from the environment enter the system, as was noted, for example, in [10].

The regularities established in the work make it possible to find the lifetime of physical, chemical, and other systems, and to model their behavior, exploring different features of their evolution and stationary states, taking into account different features of the past of thesystems.

Thermodynamics with a lifetime describes open systems far from equilibrium and can be used to study the behavior of dissipative structures and other synergetic effects. It is possible to describe the nonequilibrium behavior of arbitrary physical quantities by which the investigated physical system is open, taking into account the influence of all factors that contribute to the interaction of the system with the environment.

The introduction of thermodynamic forces $\gamma$, conjugated to the value of $\Gamma$, corresponds to the effective consideration of the influence of external influences (such as the pump parameter, etc.). The introduction of the quantities $\gamma$ and $\Gamma$ corresponds to an effective account of the openness of the system.